\documentclass[twocolumn,showpacs,aps,epsfig]{revtex4}

\usepackage{graphicx}
\usepackage{epstopdf}
\usepackage{latexsym}

\def\be{\begin{equation}}
\def\ee{\end{equation}}
\def\bea{\begin{eqnarray}}
\def\eea{\end{eqnarray}}

\newcommand{\bear}{\begin{eqnarray}}

\newcommand{\eear}{\end{eqnarray}}

\newbox\pippobox

\def\6{\partial}
\def\bx{{\bf x}}
\def\bk{{\bf k}}

\def\a{\alpha}
\def\nn{\nonumber}

\def\e{\epsilon}

\def\sq
\def\a{\alpha}

\def\l{\lambda}

\def\na{\nabla}

\def\e{\epsilon}

\begin{document}

\title{{\bf Dynamical Scalar Degree of Freedom in Ho\v rava-Lifshitz Gravity}}

\author{Rong-Gen Cai$^{1,}$\footnote{Email: cairg@itp.ac.cn}}

\author{Bin Hu$^{1,}$\footnote{Email: hubin@itp.ac.cn}}

\author{Hong-Bo Zhang$^{1,2,}$\footnote{Email:
 hbzhang@itp.ac.cn}}

\affiliation{$^{1}$Key Laboratory of Frontiers in Theoretical
Physics, Institute of Theoretical Physics, Chinese Academy of
Sciences, P.O. Box 2735, Beijing 100190, China\\
${}^{2}$Graduate University of Chinese Academy of Sciences, YuQuan
Road 19A, Beijing 100049, China}

\date{\today}

\begin{abstract}
We investigate the linear cosmological perturbations of Ho\v
rava-Lifshitz gravity in a FRW universe without any matter. Our
results show that a new gauge invariant dynamical scalar mode
emerges, due to the gauge transformation under the
``foliation-preserving" diffeomorphism and ``projectability
condition", and it can produce a scale invariant power spectrum.
In the infrared regime with $\lambda=1$, the dynamical scalar
degree of freedom turns to be a non-dynamical one at the linear
order level.
\end{abstract}


\maketitle

Recently, a power-counting renormalizable ultra-violet (UV)
complete quantum gravity theory was proposed by Ho\v
rava~\cite{hor1,hor2,hor3}. This theory is characterized by the
anisotropic scaling between time and space, so the complete
diffeomorphism invariance of General Relativity (GR) is lost,
instead the Ho\v rava-Lifshitz (HL) gravity is invariant under the
so-called ``foliation-preserving" diffeomorphism. Since this
theory was proposed, a great deal of efforts have been made,
including studies of cosmology
\cite{Greece,Calcagni:2009ar,Mukohyama:2009gg,Mukohyama:2009zs,Piao:2009ax,
Gao:2009bx,Brandenberger:2009yt,Takahashi:2009wc} and black hole
physics \cite{Cai:2009pe,Cai:2009ar,Cai:2009qs}, among others
\cite{Chen:2009ka,Kluson:2009sm,Nikolic:2009jg,Colgain:2009fe}.
Due to the differences of diffeomorphism group between HL and GR,
we expect to see some new dynamical degrees of freedom of
gravitational fields in HL gravity. Indeed, a new dynamical scalar
degree of freedom of gravitational fields is firstly discussed by
Ho\v rava~\cite{hor1} in Minkowski spacetime. The motivation of
this paper is to investigate the dynamical behavior of this scalar
mode in a cosmological background.

Let us begin with a brief review about the HL gravity~\cite{hor3}.
In order to construct a UV renormalizable quantum gravity, one
possible way is to introduce high order spatial derivative
operators, which make the graviton propagator fall off
sufficiently rapidly at large momenta. On the other hand, in order
for the theory to be unitary, the Lagrangian can only be quadratic
in first time derivatives of the spatial metric. As a consequence,
the theory has the anisotropic scaling between space and time. For
instance, in $3+1$ dimensions the coordinates $(t,\bx)$ scale as
 \be\label{scaling}  t\to
\ell^z~t\;,\qquad \bx \to \ell~\bx\;,\ee
 where $z$
is called dynamical critical exponent. In terms of ADM formalism
the metric can be written as
 \be\label{ADM} ds^2=-N^2
~dt^2+g_{ij}\Big(dx^i+N^idt\Big)\Big(dx^j+N^jdt\Big)\;,\ee
 where the spatial metric, lapse function and shift vector scale as
 \be\label{lapse shift} g_{ij} \to g_{ij}\;,\qquad N \to N\;,\qquad
 N_i \to \ell^{z-1}N_i\;.\ee

The action of the non-relativistic renormalisable gravitational
theory proposed by Ho\v rava contains two parts. The part of
kinetic term is
 \be\label{kinetic} S_K={2\over \kappa^2}\int dtd^3x
\sqrt{g}N\Big(K_{ij}K^{ij}-\l K^2\Big)\;,\ee
 where the extrinsic curvature reads
\be\label{extri-curv} K_{ij}={1\over 2N}\Big(\dot
g_{ij}-\nabla_{i}N_j-\nabla_jN_i\Big)\;.\ee The part of the
potential term in the so-called ``detailed-balance condition" can
be written down as
 \bea \label{potential} S_{V}&=&\int dt
d^3x\sqrt{g}N~\left[-{\kappa^2\over 2w^4}C_{ij}C^{ij}\right.\nonumber\\
&&+{\kappa^2\mu\over 2w^2} {\e^{ijk}\over
\sqrt{g}}R_{il}\na_j{R^l}_{k}-{\kappa^2\mu^2\over
8}R_{ij}R^{ij}\nonumber\\
&&\left.+{\kappa^2\mu^2\over 8(1-3\l)} \left({1-4\l\over
4}R^2+\Lambda~R-3\Lambda^2\right)\right]\;, \eea
 where
$\lambda$, $\kappa$, $\mu$, $w$, $\Lambda$ are coupling constants,
$\e^{ijk}$ is the antisymmetric tensor defined by $\e_{123}=1$ and
the Cotton tensor reads
 \be\label{cotton}
C^{ij}={\e^{ikl}\over \sqrt{g}}\na_k\left({R^j}_{l}-{1\over
4}R{\delta^{j}}_{l}\right)\;.\ee The Cotton tensor term in the
first line of (\ref{potential}), which scales as $z=3$, is
introduced in order for the theory to be power-counting
renormalizable in $3+1$ dimensions. The other terms in the second
and third lines will make the theory undergo a classical flow to
$z=1$ in the infrared (IR) regime, where the coupling constant
$\lambda$ would be expected to flow to $\lambda=1$. And then GR
would be expected to be recovered in the IR regime.

By combining (\ref{kinetic}) and (\ref{potential}), the full action
$S_1=S_K+S_V$ can be expressed as
 \bea
\label{full1}S_1&=&\int dtd^3x\sqrt{g}N\Big[\alpha_1
\Big(K_{ij}K^{ij}-\l K^2\Big)\nonumber\\
&&+\beta_1 C_{ij}C^{ij}+\gamma_1 {\e^{ijk}\over
\sqrt{g}}R_{il}\na_j{R^l}_{k}\nonumber\\
&&+\zeta_1 R_{ij}R^{ij}+\eta_1 R^2+\xi_1 R+\sigma_1\Big]\;,\eea
 with coupling constants
  \bea\label{coef} &&\alpha_1=
\frac{2}{\kappa^{2}}\;,\quad \beta_1=
-\frac{\kappa^{2}}{2w^{4}}\;,\quad \gamma_1=
\frac{\kappa^{2}\mu}{2w^{2}}\;,\quad\nonumber\\
&&\zeta_1=-\frac{\kappa^{2}\mu^{2}}{8}\;,\quad
 \eta_1=
\frac{\kappa^{2}\mu^{2}}{8(1\!-\!3\lambda)}\frac{1\!-\!4\lambda}{4}\;,\quad\nonumber\\
&&\xi_1= \frac{\kappa^{2}\mu^{2}}{8(1\!-\!3\lambda)}\Lambda\;,\quad
\sigma_1=
\frac{\kappa^{2}\mu^{2}}{8(1\!-\!3\lambda)}(-3\Lambda^{2})\;,\eea
 where $\mu$ and $w^2$ are real constants, and have their origin as the
Newton constant and Chern-Simons coupling of Euclideanised
three-dimensional topologically massive gravity \cite{DJT}. In order
to have a real speed of light, $\Lambda$ must have to be negative
for $\lambda>1/3$~\cite{hor3}, which leads to a negative effective
cosmological constant $\sigma_1$. This is not consistent with
current cosmological observation. To have a positive cosmological
constant, one may make an analytic continuation of those
parameters~\cite{LMP}
 \be \mu\rightarrow
i\mu\;,\qquad w^2\rightarrow -iw^2\;,\ee
 then the action
(\ref{full1}) changes to
 \bea \label{full2}S_2&=&\int
dtd^3x\sqrt{g}N\Big[\alpha_2 (K_{ij}K^{ij}-\l
K^2)\nonumber\\
&&+\beta_2 C_{ij}C^{ij}+\gamma_2 {\e^{ijk}\over
\sqrt{g}}R_{il}\na_j{R^l}_{k}\nonumber\\
&&+\zeta_2 R_{ij}R^{ij}+\eta_2 R^2+\xi_2 R+\sigma_2\Big]\;,\eea
 with coefficients
 \bea &&\alpha_2=\alpha_1\;,\quad
\beta_2=-\beta_1\;,\quad \gamma_2=-\gamma_1\;,\quad
\zeta_2=-\zeta_1\;,\nonumber\\
&&\eta_2=-\eta_1\;,\quad \xi_2=-\xi_1\;,\quad
\sigma_2=-\sigma_1\;.\eea In addition to the gravitational sector
(\ref{full1}) or (\ref{full2}), we can also add some matter sectors
to HL theory
 \be\label{matter}
S_{M}=\int d^3xdt \sqrt{g}N~{\cal L}_{\rm
matter}(N,N_i,g_{ij})\;.\ee

Before deriving the constraint and dynamical equations of the
theory, let us stress that, in order for the theory to be
tractable, the lapse function $N$ should be a function of time
coordinate $t$ only, otherwise it would lead to difficulties in
quantization, at least in the absence of extra gauge symmetries.
This prescription is known as the ``projectability
condition"~\cite{hor3,Sotiriou:2009gy}.

Because of the ``projectability condition" on the lapse function
$N(t)$, we can only obtain the spatially integrated Hamiltonian
constraint by varying the action with respect to $N(t)$
 \bea\label{hamilton} 0&=&\int d^3x \sqrt{g}\Big\{-\alpha_m\left(K_{ij}K^{ij}-\l
K^2\right)+\beta_m C_{ij}C^{ij}\nonumber\\
&&+\gamma_m {\e^{ijk}\over \sqrt{g}}R_{il}\na_j{R^l}_{k}+\zeta_m
R_{ij}R^{ij}+\eta_m R^2\nonumber\\
&&+\xi_m R+\sigma_m+J_N\Big\}\;,\eea
 with $m=1,2$ and
  \bea J_N={\cal L}_{\rm
matter}+N{\delta {\cal L}_{\rm matter}\over \delta N}\;.\eea
 The
equation of motion for $N_i$ gives momentum constraint
\be\label{momentum} 2\alpha_m(\na_jK^{ji}-\l\na^i K)+N{\delta
{\cal L}_{\rm matter}\over \delta N_i}=0\;.\ee The equation of
motion for $g_{ij}$ is very lengthy, and one can find the explicit
expression in \cite{Greece}.

Now we investigate HL gravity in a flat Friedmann-Robertson-Walker
universe
 \be ds^2=-dt^2+a^2(t)\delta_{ij}dx^idx^j\;,\ee
  with the scale factor $a(t)$. For further simplification we
turn off the matter sector (\ref{matter}). In that case, the
spatially integrated Hamiltonian constraint (\ref{hamilton}) reads
 \be\label{friedman1}H^2=\frac{\sigma_m}{3\alpha_m(1-3\lambda)}\;,\ee
 where the Hubble parameter $H=\dot a/a$, where a dot denotes the derivative
 with respect to the cosmic time $t$, while the
 momentum constraint is trivially satisfied.

The equation of motion for $g_{ij}$ gives
\be\label{friedman2}2\alpha_m(3\l-1)\left[\dot H+{3\over
2}H^2\right]=-\sigma_m\;.\ee
 By virtue of the above equations, one can easily verify that for the
action (\ref{full1}) $H^2<0$, while for the action (\ref{full2})
$H^2>0$. Namely in the latter case, a de Sitter solution exists.
Due to cosmological interest, we will consider the latter case and
omit the subscript $m=2$ in the rest of this paper.

Due to the anisotropic scaling of temporal and spatial
coordinates, the time coordinate $t$ plays a privileged role in
this theory, and the symmetry of HL theory is smaller than the one
of GR. Precisely, HL gravity is invariant under so called
``foliation-preserving" diffeomorphism, where the coordinate
transformations have the form
 \be\label{foli}t\rightarrow\tilde{t}=f(t)\;,\quad x^i\rightarrow\tilde{x}^i=h^i(t,\bx)\;,\ee
 with $f(t)$ is a function of $t$ only. This is a big difference
 from the case of GR.

In cosmological perturbation theory, the metric perturbations are
usually categorized into three distinct types: scalar, vector and
tensor perturbations
 \bea\label{so3}&&\delta  g_{00}=-2a^2\phi\;,\qquad \delta
g_{0i}=a^2\partial_iB+a^2Q_i\;,\\
&&\delta
g_{ij}=a^2h_{ij}-a^2(\partial_iW_j+\partial_jW_i)-2a^2(\psi\delta_{ij}-\partial_i\partial_jE)\;,\nonumber\eea
 where $\phi$, $\psi$, $E$, $B$ are four scalar modes, $Q_i$,
$W_i$ are two vector modes which satisfy
$\partial^iQ_i=\partial^iW_i=0$, and $h_{ij}$ is the
transverse-traceless tensor mode $h_{ij,j}=h_{ii}=0$. Under an
infinitesimal ``foliation-preserving" coordinate transformation
 \be\label{coord
trans}t\rightarrow\tilde t=t+\e^0(t)\;,\qquad x^i\rightarrow\tilde
x^i=x^i+\e^i(t,\bx)\;,\ee
 and
decomposing $\e^i$ into
 \be \e^i=\partial^i\e(t,\bx)+\zeta^i(t,\bx)\;,\ee
  with
$\partial_i\zeta^i=0$, we can obtain the transformation rules
\begin{eqnarray}
\label{dic s1}
\phi&\rightarrow&\tilde\phi=\phi-\dot{\e}^0\;,\\
\label{dic s2}
\psi&\rightarrow&\tilde\psi=\psi+H\e^0\;,\\
\label{dic s3}
B&\rightarrow&\tilde B=B-a(\frac{\epsilon}{a^2})^{\dot{}}\;,\\
\label{dic s4} E&\rightarrow&\tilde E=E-\frac{1}{a^2}\e\;,
\end{eqnarray}
for scalar modes,
\begin{eqnarray}
\label{dic v}
Q_i&\rightarrow&\tilde Q_i=Q_i-a(\frac{\zeta_i}{a^2})^{\dot{}}\;,\\
W_i&\rightarrow&\tilde W_i=W_i+\frac{1}{a^2}\zeta_i\;,
\end{eqnarray}
 for vector modes,
 and the transformation for tensor modes is the same as that in GR
 because of the ``foliation-preserving" diffeomorphism.
  Unlike what happens in GR \cite{nini}, however, we can see
from (\ref{dic s2}) that we are forbidden to choose the
``spatially flat gauge" in HL gravity since the infinitesimal
parameter $\e^0$ is the function of $t$ only. Further, due to the
``projectability condition", the lapse function can be set
globally to unity, i.e., we can gauge $\phi$ mode by choosing a
proper initial time. And the residual coordinate freedom can be
gauged by virtue of (\ref{dic s4}), thus we can completely fix the
coordinates.

With help of those transformation rules, one can easily build up
the gauge-invariant variables as
 \begin{eqnarray}
\label{gaug inv1} \Phi&=&\phi+(\frac{\psi
}{H})^{\dot{}}\;,\\
\label{gaug inv2} \Pi&\equiv&B-a\dot E\;,\\
\label{gaug inv3} \Psi&=&\psi+H \int_{t_0}^{t}\phi dt'\;,
\end{eqnarray}
 where only two of them are independent, since we have
  $\Phi=(\Psi/H)^{\dot{}}$ from (\ref{gaug inv1}) and (\ref{gaug inv3}).
In the gauge ($\phi=0$, $E=0$), $\psi$ coincides with the gauge
invariant variable $\Psi$. Thus it is very convenient to take this
gauge in discussing cosmological scalar perturbations. In this
gauge, the metric becomes
 \be\label{gaug
fix}ds^2=-dt^2+2a\partial_iBdtdx^i+a^2(1-2\psi)\delta_{ij}dx^idx^j\;.\ee
Substituting (\ref{gaug fix}) into (\ref{full2}) and performing
lots of straightforward but very lengthy calculations, we obtain
the action of the scalar perturbations
 \bea
\label{sk} S_K&=&\int dtd^3x \alpha
a^3\Big\{(1-3\lambda)\Big[6H\psi\dot\psi+3\dot\psi^2+\frac{2}{a}\dot\psi\partial^2B\nonumber\\
&&+
\frac{9}{2}H^2\psi^2\Big]+\frac{1-\lambda}{a^2}B\partial^4B\Big\}\;,\eea
 \bea\label{sv} S_V&=&\int dtd^3x\Big\{\frac{2(3\zeta+8\eta)}{a}\psi\partial^4\psi+\frac{3\sigma}{2}a^3\psi^2\nn\\
 &&-2\xi
a\psi\partial^2\psi\Big\}\;,\eea
 with
$\partial^2\equiv\delta^{ij}\partial_i\partial_j$. Although the
action (\ref{full2}) contains terms such as $C_{ij}C^{ij}$ and
$R_{il}\na_j{R^l}_{k}$, the highest order of spatial derivatives
in the action is $\partial^4$ for the sake of the antisymmetric
tensor $\e^{ijk}$ and flat universe. However, if we abandon the
``detailed-balanced condition",  $\partial^6$ terms will appear.

After dropping some surface terms and using the background equations
(\ref{friedman1}) (\ref{friedman2}), the spatially integrated
Hamiltonian constraint becomes
 \be\label{1st ham}\int d^3x~6\alpha(1-3\lambda)H\dot{\psi}=0\;.\ee
The equation of motion for $\partial_iB$ gives the momentum
constraint
 \be\label{eom b} \partial_i\Big\{(3\lambda-1)\dot{\psi}+(\lambda-1)\frac{1}{a}\partial^2B\Big\}=0\;,\ee
 and the dynamical equation of motion for the gravitational field
$\psi$ is given by
 \bea\label{eom psi} && 6\alpha(1-3\lambda)\Big[\ddot\psi+3H\dot\psi+\frac{1}{3a}\Big(\partial^2\dot
B+2H\partial^2B\Big)\Big]\nn\\
&&~~~~~~~~~~~~~~+\frac{4\xi}{a^2}\partial^2\psi-\frac{4(3\zeta+8\eta)}
{a^4}\partial^4\psi =0\;.\eea
 This equation is the main result of this paper.

$\bullet$ When $\lambda=1/3$, the first line of (\ref{eom psi})
vanishes, the coefficients in second line diverge and the
Hamiltonian constraint (\ref{1st ham}) is trivially satisfied. So,
the evolution of $\psi$ cannot be determined by the classical
equation of motion (\ref{eom psi}). This strange feature indicates
that we need to take account of the quantum Renormalization Group
flows at the UV fixed point.

$\bullet$ When $\lambda\neq 1/3$ and $1$, substituting the
momentum constraint (\ref{eom b}) into the dynamical equation
(\ref{eom psi}), we obtain
 \be\label{eom
dyna}\ddot\psi+3H\dot\psi+\frac{(1-\lambda)\xi}{\alpha(1-3\lambda)
a^2}\partial^2\psi-\frac{(1-\lambda)(3\zeta+8\eta)}{\alpha(1-3\lambda)
a^4}\partial^4\psi=0\;.\ee

$\bullet$ When $\lambda=1$, the momentum
 constraint (\ref{eom b}) becomes
 \be\label{red mom}
 \partial_i\dot\psi=0\;,\ee
 with the general solution
 \be\label{gen sol}
 \psi(t,\bx)=f(t)+g(\bx)\;.\ee
  Furthermore, we can obtain $f(t)={\rm const.}$ by plugging this solution into the
  spatially integrated Hamiltonian constraint (\ref{1st ham}). As a result, $\psi$ is independent of
  time and thus $\psi$ is not a dynamical degree of freedom at least the linear order level.
  Of course, it does not exclude $\psi$ is a dynamical one once
  higher order perturbations are taken into account.

During the derivation of the above dynamical equations, we have
not considered the spatially integrated Hamiltonian constraint
(\ref{1st ham}). Now we explain why we can safely ignore this
constraint. It turns out to be more convenient to discuss this
issue in Fourier space
 \be\label{fouier}\psi(t,\bx)=\frac{1}{(2\pi)^3}\int d^3\bk
~\psi_k(t)e^{i\bk\cdot\bx}\;,\ee
 then (\ref{1st ham}) becomes
 \bea\label{f1st ham}&&6\alpha(1-3\lambda)\dot{\psi}_{k=0}=0\;.\eea
 we can see from (\ref{f1st ham}) that the spatially integrated
Hamiltonian constraint only constrains $k=0$ mode, while this mode
can be absorbed into the background. As a result, we only need to
consider the momentum constraint (\ref{eom b}) in HL gravity.

In what follows we will focus on solving the dynamical equation
(\ref{eom dyna}) for the case $\lambda\neq 1/3$ and $1$. After
introducing a new field $\psi\equiv\chi/a$, (\ref{eom dyna}) can
be written in Fourier space as
 \bea\label{eom dyna2}
 && \chi_k^{''}(\tau)+\Big[-\frac{(1-\lambda)^2c^2H^2}{(3\lambda-1)\Lambda}k^4\tau^2\nn\\
&&~~~~~~~~~~~+\frac{(1-\lambda)c^2}{3\lambda-1}k^2-\frac{2}{\tau^2}\Big]\chi_k(\tau)=0\;,\eea
 where prime stands for derivative with respect to
conformal time ($d\tau=dt/a$) and $c$ for the speed of light
$c\equiv\sqrt{\xi/\alpha}=\sqrt{\kappa^4\mu^2\Lambda/16(3\lambda-1)}$.
The requirement that the speed of light be real implies that
$\Lambda$ be positive for $\lambda>1/3$. In the rest of this paper
we will concentrate on this case.

Comparing the three terms in the square bracket, we can divide our
discussions into three cases.

$\bullet$ Case $1$: when $k^2\tau^2\gg T_1$
($T_1\equiv\Lambda/H^2$) and $k^2\tau^2\gg T_2$
($T_2\equiv\sqrt{\Lambda/c^2H^2}$), $k^4$ term dominates. In this
case, (\ref{eom dyna2}) reduces to
 \be\label{k4 domi}
\chi_k^{''}(\tau)-\frac{(1-\lambda)^2c^2H^2}{(3\lambda-1)\Lambda}
k^4\tau^2\chi_k(\tau)=0\;.
 \ee
This equation has a general solution with form
  \bea\label{sol k4 domi}
   \chi_k(\tau)&=&C_1{\rm
ParabolicCylinderD}\left[-\frac{1}{2},
\sqrt{2}\omega^{1/4}k\tau\right]\nn\\
&+&C_2{\rm ParabolicCylinderD}\left[-\frac{1}{2},
i\sqrt{2}\omega^{1/4}k\tau\right]\;\eea
 where $C_1, C_2$ are two
integration constants,
$\omega=(1-\lambda)^2c^2H^2/(3\lambda-1)\Lambda$ and ${\rm
ParabolicCylinderD}[\alpha, x]$ is the parabolic cylinder
function. It is easy to check that both terms in (\ref{sol k4
domi})are exponentially decaying modes in the region
$-\infty<\tau<0$. In order to see the behavior of this solution
more clearly, we can use the WKB approximation to solve (\ref{k4
domi}) in the large $k$ limit. Assuming a trial solution in the
form of an asymptotic series expansion
 \be\label{trial}\chi_k(\tau)=\exp\left\{\frac{1}{\epsilon}
\sum_{n=0}^{\infty}\epsilon^nS_n(\tau)\right\}\;,\ee
 and plugging it into (\ref{k4 domi}),
 at the leading order, the solution
reads
  \be\label{sol k4 domi3}
 \chi_k(\tau)=\exp\left\{k^2C_3+k^2\frac{\omega^{1/2}}{2}\tau^2\right\}\;,\ee
 where $C_3$ is an integration
constant and the second term in the exponent makes the solution
 decay exponentially when the conformal time
$\tau$ evolves from $-\infty$ to $0$, which is consistent with the
asymptotic behavior of the solution (\ref{sol k4 domi}).

$\bullet$ Case $2$: when $k^2\tau^2\gg T_3$ ($T_3\equiv1/c^2$) and
$k^2\tau^2\ll T_1$, $k^2$ term will dominate  over other two terms
 \be\label{k2 domi}\chi_k''(\tau)+c_s^2k^2\chi_k(\tau)=0\;,\ee
  where
the sound speed $c_s^2\equiv(1-\lambda)c^2/(3\lambda-1)$ and a
real sound speed requires $1/3<\lambda<1$. The solution of
(\ref{k2 domi}) is a plane wave solution
 \be\label{sol k2
domi}\chi_k(\tau)=\frac{\exp\{-ikc_s\tau\}}{\sqrt{2kc_s}}\;.\ee

$\bullet$ Case $3$: when $k^2\tau^2\ll T_3$ and $k^2\tau^2\ll T_2$,
$k^0$ term is dominant. In this case, one has
 \be\label{k0
domi}\chi_k''(\tau)-\frac{2}{\tau^2}\chi_k(\tau)=0\;,\ee
 with
the solution
 \be\label{sol k0
domi}\chi_k(\tau)=C_4\tau^2+\frac{C_5}{\tau}\;,\ee
 where $C_4$,
$C_5$ are two integration constants. We can see that, the $C_4$
term is a decaying mode and the $C_5$ term is a growing one. For
simplicity we neglect the decaying mode by simply setting $C_4=0$,
then we are able to determine the absolute value of $C_5$ by
matching the absolute value of (\ref{sol k2 domi}) with it when
this mode crosses the sound horizon ($k/aH=-k\tau=1/c_s$)
\be\label{matchz} |C_5|=\frac{1}{\sqrt{2k^3c_s^3}}\;,\ee
 thus $\psi$ will be frozen on the
superhorizon scales
 \be\label{squeezed}|\psi_k|=\frac{H}{\sqrt{2k^3c_s^3}}\;.\ee
 Finally, by the
definition of scalar power spectrum $\mathcal{P}_{\psi}(k)\equiv
k^3|\psi|^2/2\pi^2$, we obtain a scale invariant spectrum
 \be\label{scale inv}\mathcal{P}_{\psi}=\frac{H^2}{4\pi^2c_s^3}\;.\ee
 We can understand these results in the following way. On the very
small scale (case $1$), the fluctuations decay exponentially; on
the relatively large scale but still deeply inside the sound
horizon (case $2$), the fluctuations oscillate until they crosse
the sound horizon; after they crosse the sound horizon (case $3$),
$k^0$ term will dominate over the other two terms and the
fluctuations $\chi$ will grow. However, this growth is compensated
by the growth of scale factor, consequently $\psi$ is frozen on
the superhorizon scale.

In conclusion, in this paper we investigated the linear cosmological
perturbations of HL gravity without any matter in a flat FRW
universe. We studied the gauge transformation under the
``foliation-preserving" diffeomorphism, derived the rigorous
equation of motion for scalar perturbations and discussed the
dynamical behavior of this mode. Our results showed that this mode
evolved dynamically with time when $1/3<\lambda<1$ and could produce
a scale invariant spectrum. In the regime with $\lambda=1$, where GR
was expected to be recovered, the dynamical scalar mode became a
non-dynamical one.  It is of great interest to investigate whether
this conclusion keeps valid beyond the linear perturbations and some
matter sectors are included.


\begin{acknowledgments}
We thank Yi Ling, Yan Liu, Ya-Wen Sun, Anzhong Wang and Jun'ichi
Yokoyama for a lot of valuable discussions. This work was
supported in part by the Chinese Academy of Sciences with Grant
No. KJCX3-SYW-N2 and the NSFC with Grant No. 10821504 and No.
10525060.
\end{acknowledgments}


\vspace*{0.2cm}
\begin{thebibliography}{99}

\bibitem{hor1}
  P.~Horava,
  JHEP {\bf 0903}, 020 (2009)
  [arXiv:0812.4287 [hep-th]].

\bibitem{hor2}
  P.~Horava,
  arXiv:0811.2217 [hep-th].

\bibitem{hor3}
  P.~Horava,
  Phys.\ Rev.\  D {\bf 79}, 084008 (2009)
  [arXiv:0901.3775 [hep-th]].

\bibitem{DJT}
  S. Deser, R. Jackiw and S. Templeton,
  Annals Phys.\  {\bf 140}, 372 (1982)
  [Erratum-ibid.\  {\bf 185}, 406.1988\ APNYA,281,409
  (1988\ APNYA,281,409-449.2000)].

\bibitem{nini}
  M.~Giovannini,
  Int.\ J.\ Mod.\ Phys.\  D {\bf 14}, 363 (2005)
  [arXiv:astro-ph/0412601].

\bibitem{LMP}
  H.~Lu, J.~Mei and C.~N.~Pope,
  arXiv:0904.1595 [hep-th].

\bibitem{Sotiriou:2009gy}
  T.~P.~Sotiriou, M.~Visser and S.~Weinfurtner,
  arXiv:0904.4464 [hep-th].

\bibitem{Greece}
  E.~Kiritsis and G.~Kofinas,
  arXiv:0904.1334 [hep-th].

\bibitem{Calcagni:2009ar}
  G.~Calcagni,
  arXiv:0904.0829 [hep-th].

\bibitem{Mukohyama:2009gg}
  S.~Mukohyama,
  arXiv:0904.2190 [hep-th].

\bibitem{Mukohyama:2009zs}
  S.~Mukohyama, K.~Nakayama, F.~Takahashi and S.~Yokoyama,
  arXiv:0905.0055 [hep-th].

\bibitem{Piao:2009ax}
  Y.~S.~Piao,
  arXiv:0904.4117 [hep-th].

\bibitem{Gao:2009bx}
  X.~Gao,
  arXiv:0904.4187 [hep-th].

\bibitem{Brandenberger:2009yt}
  R.~Brandenberger,
  arXiv:0904.2835 [hep-th].

\bibitem{Takahashi:2009wc}
  T.~Takahashi and J.~Soda,
  arXiv:0904.0554 [hep-th].

\bibitem{Cai:2009pe}
  R.~G.~Cai, L.~M.~Cao and N.~Ohta,
  arXiv:0904.3670 [hep-th].

\bibitem{Cai:2009ar}
  R.~G.~Cai, Y.~Liu and Y.~W.~Sun,
  arXiv:0904.4104 [hep-th].

\bibitem{Cai:2009qs}
  R.~G.~Cai, L.~M.~Cao and N.~Ohta,
  arXiv:0905.0751 [hep-th].

\bibitem{Chen:2009ka}
  B.~Chen and Q.~G.~Huang,
  arXiv:0904.4565 [hep-th].

\bibitem{Kluson:2009sm}
  J.~Kluson,
  arXiv:0904.1343 [hep-th].

\bibitem{Nikolic:2009jg}
  H.~Nikolic,
  arXiv:0904.3412 [hep-th].

\bibitem{Colgain:2009fe}
  E.~O.~Colgain and H.~Yavartanoo,
  arXiv:0904.4357 [hep-th].


\end{thebibliography}
\end{document}